\documentclass[twocolumn]{aastex62}

\newcommand{\mhi}{\ensuremath{M_{HI}}}

\submitjournal{ApJ}
\shorttitle{Star Formation History of Antlia B}
\shortauthors{Hargis et al.}

\begin{document}

\title{{\it Hubble Space Telescope} Imaging of Antlia~B: \\Star Formation History and a New Tip of the Red Giant Branch Distance}

\correspondingauthor{J.R. Hargis}
\email{jhargis@stsci.edu}

\author[0000-0002-8722-9806]{J.R. Hargis}
\affiliation{Space Telescope Science Institute, 3800 San Martin Drive, Baltimore, MD, 21208, USA}

\author[0000-0001-5496-2668]{S. Albers}
\affiliation{University of California, Berkeley, Department of Astronomy, 501 Campbell Hall \#3411, Berkeley, CA 94720-3411,USA}

\author[0000-0002-1763-4128]{D. Crnojevi\'c}
\affil{University of Tampa, 401 West Kennedy Boulevard, Tampa, FL 33606, USA}
\affil{Department of Physics \& Astronomy, Texas Tech University, Box 41051, Lubbock, TX 79409-1051, USA}

\author[0000-0003-4102-380X]{D. J. Sand}
\affiliation{Steward Observatory, University of Arizona, 933 North Cherry Avenue, Rm. N204, Tucson, AZ 85721-0065, USA}

\author{D.~R. Weisz}
\affiliation{University of California, Berkeley, Department of Astronomy, 501 Campbell Hall \#3411, Berkeley, CA 94720-3411,USA}

\author[0000-0002-3936-9628]{J. L. Carlin}
\affil{LSST, 950 North Cherry Avenue, Tucson, AZ 85719, USA} 

\author{K. Spekkens}
\affiliation{Department of Physics and Space Science, Royal Military College of Canada P.O. Box 17000, Station Forces Kingston, ON K7K 7B4, Canada}
\affiliation{Department of Physics, Engineering Physics and Astronomy, Queen’s University, Kingston, ON K7L 3N6, Canada}

\author{B. Willman}
\affiliation{Steward Observatory, University of Arizona, 933 North Cherry Avenue, Rm. N204, Tucson, AZ 85721-0065, USA}
\affiliation{National Center for Optical and Infrared Astronomy, 933 North Cherry Avenue, Tucson, AZ 85721, USA}

\author[0000-0002-8040-6785]{A. H. G. Peter} 
\affiliation{CCAPP, Department of Physics, and Department of Astronomy, The Ohio State University, Columbus, OH 43210, USA}

\author{C.~J. Grillmair}
\affiliation{IPAC, California Institute of Technology, Pasadena, CA 91125, USA}

\author{A.~E. Dolphin}
\affiliation{Raytheon Company, Tucson, AZ, 85734, USA}

\begin{abstract}
A census of the satellite population around dwarf galaxy primary hosts in environments outside the Local Group is essential to understanding $\Lambda {\rm CDM}$ galaxy formation and evolution on the smallest scales.  We present deep optical {\it Hubble Space Telescope} imaging of the gas-rich, faint dwarf galaxy Antlia~B ($M_V = -9.4$) -- a likely satellite of NGC~3109 ($D = 1.3$ Mpc) -- discovered as part of our ongoing survey of primary host galaxies similar to the Magellanic Clouds.  We derive a new tip of the red giant branch (TRGB) distance of $D = 1.35 \pm 0.06$ Mpc ($m-M = 25.65 \pm 0.10$), consistent with membership in the nearby NGC~3109 dwarf association.  The color-magnitude diagram shows both a prominent old, metal-poor stellar component and confirms a small population of young, blue stars with ages $\lesssim 1$ Gyr.  We use the color magnitude diagram fitting algorithm {\tt MATCH} to derive the star formation history and find that it is consistent with the typical dwarf irregular or transitional dwarf galaxy (dTrans) in the Local Group.  Antlia~B shows relatively constant stellar mass growth for the first $\sim 10-11$ Gyr and almost no growth in the last $\sim 2-3$ Gyr.  Despite being gas-rich, Antlia~B shows no evidence of active star formation (i.e., no H$\alpha$ emission) and should therefore be classified as a dTrans dwarf.  Both Antlia~B and the Antlia dwarf (dTrans) are likely satellites of NGC~3109 suggesting that the cessation of ongoing star formation in these galaxies may be environmentally driven.  Future work studying the gas kinematics and distribution in Antlia~B will explore this scenario in greater detail.  Our work highlights the fact that detailed studies of nearby dwarf galaxies in a variety of environments may continue to shed light on the processes that drive the star formation history and evolution of dwarf galaxies more generally.  

\end{abstract}
 
\section{Introduction}\label{sec:intro}

The $\Lambda$ Cold Dark Matter ($\Lambda$CDM) model for structure formation is extremely successful in describing the Universe on large scales ($\gtrsim$10 Mpc), but continues to face challenges on smaller, subgalactic scales \citep[see][for a recent review]{Bullock17} where several ``problems" with the faint end of the galaxy luminosity function manifest themselves.  Work on both theoretical \citep[e.g.,][]{Brooks13,Sawala16,Wetzel16,Kim18} and observational \citep[e.g.,][]{Torrealba18,Koposov18} fronts have primarily focused on reconciling $\Lambda$CDM issues in the context of the Milky Way and its satellite system.  However, to truly test the $\Lambda$CDM model for structure formation on the smallest scales, observational studies of satellite populations beyond the Local Group are necessary and must sample primary halos with a range of masses, morphologies and environments.  This work is advancing with a primary focus on Milky Way-like galaxies in the Local Volume \citep[e.g.][]{chiboucas09,Sand14,Sand15a,crnojevic14b,crnojevic16,carlin16, Carlin19, toloba16,bennet17,Carrillo17,Danieli17,Smercina17,Geha17,Smercina18,Crnojevic19,Bennet19}.  

One opportunity to explore $\Lambda$CDM on smaller scales is to survey the satellite population around low mass host galaxies, similar to the Magellanic Clouds, which recent work has suggested has its own satellite system \citep[e.g.][]{Sales17, Kalli18}.  Systematic searches of this kind would not only shed light on the local LMC/SMC system, but may also help tease out the role that environment (e.g., ram-pressure or tidal stripping) and primary host galaxy mass plays in shaping a satellite system \citep[e.g.][]{Gatto13,Dooley17}.  Our survey program has published initial results for two systems: NGC~3109 (D=1.3 Mpc; $M_{*}$$\approx$7$\times$10$^8$ M$_{\odot}$) and NGC~2403 (D=3.2 Mpc; $M_{*}$$\approx$7$\times$10$^9$ M$_{\odot}$). Each search turned up new, faint dwarf galaxies -- Antlia B  around NGC~3109 \citep{Sand15a} and NGC~2403-Dw1 around NGC~2403 \citep{carlin16}.

Utilizing galaxy satellite populations as probes of small scale cosmological structure requires not only discovering new satellites but also understanding galaxy formation and evolution in the dwarf galaxy regime; that is, how baryons populate dark matter halos at small scales and how physical processes shape the present-day, observed properties of dwarf galaxies.  Studies of the resolved stellar populations of dwarf galaxies -- via color-magnitude diagrams (CMDs) and star formation histories (SFHs) -- have been essential observational tools for understanding dwarf galaxy evolution \citep[e.g.,][and references therein]{Mateo98, tolstoy09, Weisz11}.  In the last decade, a combination of deep optical CMDs from the {\it Hubble Space Telescope} ({\it HST}) and increasingly sophisticated stellar evolution models \citep[e.g.,][]{Dotter08, Girardi10, Vandenberg14, Choi16, Marigo17} have provided a systematic census of the SFHs of dwarf galaxies within $\sim 3$ Mpc \citep[e.g.,][]{McQuinn10, Weisz11, Weisz14, McQuinn15a, Skillman17}.

In this paper, we focus on deep optical $HST$ observations of Antlia~B to follow-up on the ground-based discovery of this gas-rich, faint dwarf galaxy at $D$=1.3 Mpc.  For reference, we list many of the properties of Antlia B in Table~\ref{table:properties}, including position, half-light radius ($r_{\rm half}$), absolute magnitude ($M_V$) and single-dish \ion{H}{1} gas properties. Most of these properties were derived in \citet{Sand15a}, and are used in the current work because the incomplete spatial coverage of $HST$ does not allow us to significantly update these parameters.  In Section~\ref{sec:datareduce} we discuss the observations and data reduction, and in Section~\ref{sec:cmd} we present the color-magnitude diagram (CMD).  We present an updated tip of the red giant branch (TRGB) distance to Antlia~B in Section~\ref{sec:distance}, as well as a quantitative star formation history (SFH) in Section~\ref{sec:sfh}.  We conclude the paper by putting Antlia~B in context with the other dwarf galaxies in the NGC~3109 association -- NGC~3109 itself, Antlia, Sextans A, Sextans B, and Leo~P -- as well as that of the Local Group (Section~\ref{sec:discussion}).

\section{{\it HST} Observations} \label{sec:datareduce}

\begin{figure*}
\begin{center}
\includegraphics{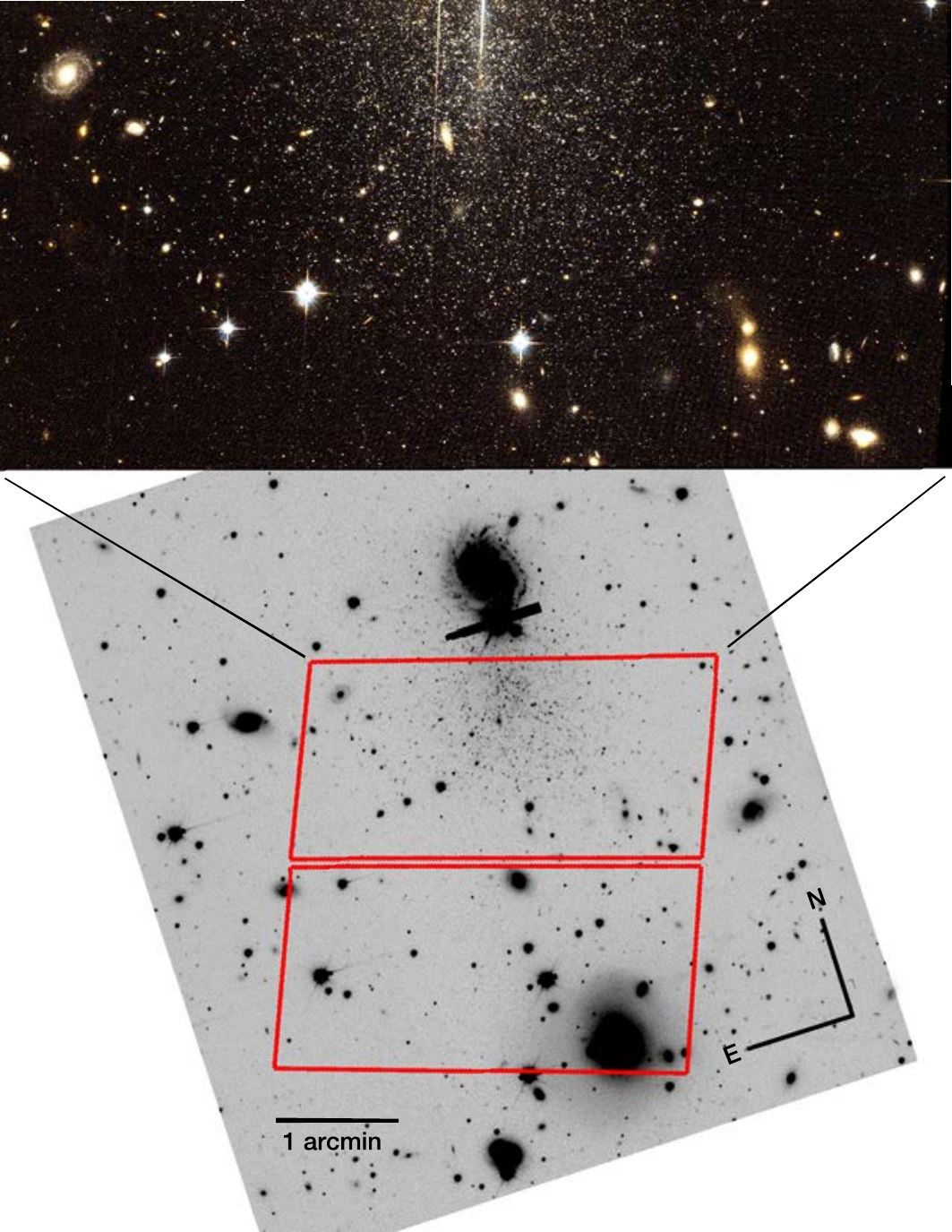}
\caption{
{\bf Top:} Color composite of Antlia~B constructed from the northernmost {\it HST/ACS} chip using the F606W and F814W images.  For scale, 1 arcminute = 393 pc at the distance of Antlia~B ($D = 1.35~\rm{Mpc}$; see $\S\ref{sec:distance}$).  {\bf Bottom:} DECam $r$-band stacked image of Antlia~B \citep{Sand15a} shown with the {\it HST/ACS} footprint ({\it red box}) of the imaging presented in this study.  The {\it HST/ACS} pointing and position angle were chosen to maximize the area coverage of Antlia B while minimizing contamination from the bright foreground star and the background galaxy north of Antlia B.  The {\it HST} data clearly show the stellar overdensity that is Antlia B.\label{fig:dwarf}}
\end{center}
\end{figure*}

Observations of Antlia~B were taken under {\it HST} program HST-GO-14078 (PI: J. Hargis) on 2017 January 4 with the Advanced Camera for Surveys \citep[ACS,][]{ACS}, using the Wide Field Channel.  The observations were taken in the F606W and F814W filters, with exposure times of 934 and 1142 s, respectively.  A two-point dither was taken for each filter to help with cosmic ray and hot pixel removal.  A color composite of one of the ACS chips is shown in Figure~\ref{fig:dwarf} overlaid on a cutout of the discovery DECam image \citep{Sand15a}.  The {\it HST} panel clearly shows the over-density of stars associated with Antlia~B.  The ACS field of view was oriented so that the bright star to the North of Antlia~B was off the chip; inevitably, this means that some Antlia~B stars were not on the ACS chip, and bleed trails from the bright star affect some stellar photometry at the edge of the field of view. This does not affect our main science goals to measure the distance and SFH of the dwarf.  

The imaging data were reduced using the CALACS pipeline (Version 8.3.5) and retrieved from the Mikulski Archive for Space Telescopes (MAST).  All the {\it HST} data used in this paper can be found in MAST here: \dataset[https://doi.org/10.17909/t9-ata8-2294]{https://doi.org/10.17909/t9-ata8-2294}.  Point spread function (PSF) photometry was performed on the CTE-corrected {\tt .flc} images with the \texttt{DOLPHOT} v2.0 photometry package, a version of \texttt{HSTPHOT} \citep[][]{Dolphin00} that has been modified for use with ACS.  We reduced the data using the \texttt{DOLPHOT} parameters and pre- and post-processing steps prescribed in \citet{Williams14}.  To construct our final list of good stars, we culled the raw photometric catalogs keeping only sources that passed the following measurement criteria: $crowd_{F606W} + crowd_{F814W} <1$, $sharp_{F606W}^2 + sharp_{F814W}^2 <0.1$, $SNR_{F606W} > 5$ and $SNR_{F814W} > 5$.  There is no evidence for crowding, even at the center of Antlia B.

We performed artificial star tests (ASTs) in order to quantify the photometric errors and incompleteness in our observations. A total of 500,000 ASTS, implanted one star at a time, were distributed uniformly both in color-magnitude space (i.e., across the relevant region of the CMD) and spatially across the field of view so as to avoid crowding.  ASTs were injected up to 2 magnitudes fainter than the faintest detected stars in order to adequately sample regions of low completeness.  Photometry and quality cuts were performed in an identical manner to those performed on the original photometry.  Photometric errors are shown as a function of F814W magnitude at the approximate color of the Antlia B ridge line in Figure~\ref{fig:cmd}. We are 50\% (90\%) complete in F814W at $\sim${26.71 (26.31)} and in F606W at $\sim${ 27.40 (27.00)} mag.

We present our final photometric catalog of resolved stars in Table~\ref{tab:photometry}.  The calibrated F814W and F606W magnitudes list in the table are uncorrected for Galactic extinction.  In addition to the full {\tt DOLPHOT} output, we also include F606W and F814W Milky Way extinction values on a star by star basis, using the dust maps of \citet{Schlegel98} and coefficients of \citet{Schlafly11}.  All CMDs presented in this work have been extinction-corrected with these values.  For reference, Antlia~B has a typical color excess of $E(B-V) \approx 0.080$ mag.

\section{Color Magnitude Diagram and Stellar Population Spatial Variations}\label{sec:cmd}

The CMD within the half-light radius ($r_{\rm half}$=273 pc or 43.2") of Antlia~B is shown in Figure~\ref{fig:cmd}a.  Antlia~B has both a significant old, metal-poor red giant branch (RGB) population and a prominent intermediate age, more metal-rich red clump (RC) population.  For comparison, in Figure~\ref{fig:cmd}b we show theoretical isochrones for old, metal-poor populations (age = 13.5 Gyr; [Fe/H] = $-$1.5, $-$2) from \citet{Dotter08} and young, more metal-rich populations (ages $250, 400, 700$ Myr; [Fe/H] = $-$1) from \citet{Marigo17}.  Antlia~B shows evidence of a small population of younger, blue stars at F606W$-$F814W $\approx 0$ consistent with an  age of $< 1$ Gyr (see blue selection box in Figure~\ref{fig:cmd}b-\ref{fig:cmd}d).  We find no evidence for very recent star formation (ages $\lesssim 10$ Myr), consistent with the ground-based H$\alpha$ imaging from \citet{Sand15a}.

To explore the spatial variations in the stellar populations of Antlia~B, we define three spatial regions of equivalent area: one within $r_{\rm half}$, one just outside of the half-light radius ($r_{\rm half} < r < 1.6~r_{\rm half}$), and one in a representative ``background" region ($r \sim 3~r_{\rm half}$).  The background region was chosen to avoid the contamination stemming from the over-density of sources around a bright background galaxy cluster (Abell S0620A) to the south-west of Antlia~B.  Figure~\ref{fig:props} shows the regions overlayed on the $HST$ field-of-view and the corresponding CMDs are shown in Figure~\ref{fig:cmd}c-\ref{fig:cmd}d.  The young, blue stellar component is clearly spatially concentrated within the inner $r_{\rm half}$.  Neither the outer nor the background region shows evidence of a younger stellar population.  This suggests that the younger, more metal-rich population within $r_{\rm half}$ is not the result of contamination.  The CMD of the background region shows structure consistent with the CMD of the central regions of Antlia~B.  Sampling additional regions at larger galactocentric radii show that while the surface density of stars is lower, there is clear evidence for Antlia~B stars out to $\sim 3~r_{\rm half}$.

A concentrated spatial distribution of younger stars or star forming regions relative to an extended distribution of older stars is typical of dwarf galaxies in the Local Volume \citep[see][and references therein]{Stinson09}.  For galaxies in the NGC~3109 dwarf association in particular the oldest stellar components (age $\gtrsim 5$ Gyr) show smooth, extended spatial distributions in contrast to more recent star formation (age $\lesssim 1$ Gyr).  Sextans~A has a patchy distribution of young/intermediate age stellar populations ($\sim 50-700$ Myr) but has a smooth extended spatial component as traced by the old RGB population \citep{vanDyk98,Dohm-Palmer02,Dolphin03,Bellazzini14}.  Sextans~B shows a similar spatial structure between young and old populations, albeit with a smaller rate of current star formation \citep{Weisz11,Bellazzini14}.  NGC~3109 also has centrally concentrated regions of young blue stars (ages $\lesssim 1$ Gyr) and a spatially extended old RGB population \citep{Weisz11,Minniti99,Hidalgo08}.  The younger population of stars in the Antlia dwarf are also centrally concentrated relative to the extended population of older, metal-poor stars \citep{Penny12}.  Lastly, Leo~P has at least one active region of star formation (e.g., single O-star and \ion{H}{2} region; \citealt{Skillman13}, \citealt{Evans19}) in the central region of the galaxy, while the older RGB population shows a larger spatial extent \citep{McQuinn15b}.

\section{TRGB Distance}\label{sec:distance}

The TRGB magnitude value is an excellent distance indicator for nearby galaxies resolved into stars \citep[e.g.,][]{Lee93, Sakai97, Makarov06, Rizzi07}.  \citet{Sand15a} used the $r$-band DECam imaging of Antlia~B to obtain a TRGB distance of $D = 1.29 \pm 0.10$ Mpc ($m - M_0 = 25.56 \pm 0.16$ mag).  We redetermine the TRGB distance here with the {\it HST/ACS} dataset, which has both higher signal to noise and superior star-galaxy separation than the ground-based data.

We adopt the TRGB absolute magnitude calibration in the F814W filter from \cite{jang17}: 

\begin{equation}
\begin{array}{lr}
M^{\rm TRGB}_{F814W}= -4.015(\pm0.056)-0.159(\pm0.01) \\
\times[(F606W-F814W)_0-1.1]^2 +0.047(\pm0.02) \\
\times[(F606W-F814W)_0-1.1]
\end{array}
\end{equation}

We apply the metallicity-dependent color correction term to our photometry to obtain a more well defined measure of the TRGB \citep[see, e.g.,][]{Madore09, McQuinn16}.  We adopt the approach of \cite{Makarov06} to find the TRGB, where a pre-defined luminosity function (LF) is compared to the observed RGB LF.  The model LF has the form

\begin{equation}
 \psi = \left\{ \begin{array}{lr}
                 10^{a(m-m_{\rm TRGB})+b}, & m - m_{\rm TRGB} \ge 0, \\
                 10^{c(m-m_{\rm TRGB})},   & m - m_{\rm TRGB} < 0
                \end{array} \right.
\end{equation}

\noindent where $a$ and $c$ are the slopes of the RGB and AGB, respectively, and $b$ represents the discontinuity at the TRGB magnitude.  The model LF is convolved with the photometric uncertainty, bias, and completeness function derived from the artificial star tests, and subsequently fit with a non-linear least squares (Levenberg--Marquardt) method for increased computational speed.  As an initial guess for the algorithm, we estimated $m_{\rm TRGB}$ using the results of a Sobel edge-detection filter (see \citealt{Sakai97, Crnojevic19} for details).  In general we find that the model-fitting TRGB method provides a more robust distance estimate and smaller uncertainties than the Sobel filter, primarily because the latter method is sensitive to the choice of the LF bin size.

We derive a value of $m_{TRGB}=21.63 \pm 0.08$, corresponding to a TRGB distance of $D = 1.35 \pm 0.06$~Mpc ($(m-M)_{\rm 0} = 25.65 \pm 0.10$~mag). This is $\sim 0.05$ Mpc more distant than the ground-based TRGB distance from \citet{Sand15a}, consistent with their result within the uncertainties.  Considering the distance and projected separation ($D_{\rm proj}$=73 kpc) of Antlia~B from NGC~3109 ($D_{\rm TRGB} = 1.28 \pm 0.03$ Mpc; \citealt{Dalcanton09}), it is clear that Antlia~B is associated with NGC~3109 -- either as a bound satellite or member of the broader NGC~3109 dwarf association.  Given the distance uncertainties, whether or not Antlia~B lies within the virial radius of NGC~3109 remains an open question.  As discussed by \citet{Sand15a}, if Antlia~B lies within the virial volume of NGC~3109 ($\sim 100$ kpc) one might expect that ram-pressure stripping (or other physical mechanisms) may have removed the gas from Antlia~B (see additional discussion in Section~\ref{sec:discussion}).  

\section{Star Formation History}\label{sec:sfh}

We measure the quantitative SFH of Antlia~B using \texttt{MATCH} \citep{Dolphin02} following implementations detailed in the literature \citep[e.g.,][]{Weisz11, Weisz14}.  Here, we briefly summarize.  

\texttt{MATCH} generates a model CMD given specified parameters including IMF slope, binary fraction, distance, extinction, age and metallicity bin widths, and a given set of stellar models.  It constructs a composite model CMD, which is then convolved with the error distribution and completeness function measured from artificial star tests.  A foreground component is added to create a final model CMD.  This model CMD is compared to the observed CMD using a Poisson likelihood function. The code computes multiple realizations of the SFH (i.e., by varying weights on each age and metallicity bin) and re-evaluates the likelihood function until a maximum likelihood solution is found.

In the case of Antlia~B, we adopted parameters identical to those used in \citet{Weisz14}: a Kroupa IMF \citep{kroupa01}, a binary fraction of 0.35 with a uniform mass ratio, the Padova stellar evolution models \citep{Girardi10}, a metallicity grid ranging from $-2.3 \le [M/H] \le -0.1$ with a resolution of 0.1 and an age grid of $\log(t)=10.15-9.00$ in steps of $\Delta \log(t) = 0.05$ and $\log(t)=9.00-6.60$ in steps of $\Delta \log(t) = 0.05$.  We adopt the Tip of the Red Giant Branch (TRGB) distance as measured in \S \ref{sec:distance} and the Milky Way foreground extinction values from \citet{Schlafly11} at the position of Antlia B. Finally, we require that the mean metallicity increase monotonically with time, with an allowed scatter. We use this age-metallicity prior because SFHs measured from CMDs that do not reach below the oldest main sequence turnoff suffer from a strong age-metallicity degeneracy \citep[e.g.,][]{Weisz11}. We compute random and systematic uncertainties on the SFH as described in \citet{Weisz14}.  Random uncertainties, which are due to the finite number of stars and S/N of the CMD, are computed using a Hamiltonian Monte Carlo algorithm as described in \cite{Dolphin13}.  Here we ran the chain for $10^4$ realizations and use the 68\% confidence interval around the best fit to represent the random uncertainties.  Systematic uncertainties, which estimate the sensitivity of the SFH to the choice in underlying stellar models, are computed using 50 Monte Carlo realizations as described in \citet{Dolphin12} and \citet{Weisz14}.  

The derived SFH is shown in Figure~\ref{fig:all_sfh} and listed in Table~\ref{tab:sfh}.  Antlia~B shows a SFH consistent with the typical dwarf irregular galaxy in the Local Group \citep[e.g.,][]{Weisz11,Weisz14}.  The results show a relatively constant growth in mass for the first $\sim 10$ Gyr with the last significant rise in star formation occurring $\sim 2-3$ Gyr ago.  We discuss the SFH in the context of the other galaxies in the NGC~3109 association in Section~\ref{sec:discussion}.

\section{Discussion: Comparison to the NGC~3109 Dwarf Association}\label{sec:discussion}

We compare the derived SFH of Antlia~B to the other possible members of the NGC~3109 association and to the dwarf galaxy population in the Local Group. All galaxies have had SFHs determined from {\it HST} imaging using {\tt MATCH}, providing a comparison which minimizes systematic uncertainties.

The population of dwarf galaxies in the Local Group can be classified morphologically into two broad classes \citep[see][and references therein]{Mateo98, Weisz11}: dwarf irregulars (dIs)/dwarf spirals (dSpirals) and dwarf spheroidals (dSphs)/ellipticals (dEs).  The first class shows evidence of recent or ongoing star formation and a significant gas reservoir, while the latter shows smooth spatial distributions of stars, no recent star formation, and no significant gas mass.

Studies of nearby dwarf populations have also revealed a small but distinct third morphological class: transition dwarfs (dTrans), which show a high gas fraction like dIs but very little or no recent star formation \citep[][]{Grebel03}.  This lack of recent star formation is often characterized by a lack of H$\alpha$ emission \citep{Mateo98}.  The origin of this subclass of dwarf galaxies is unclear, but two basic scenarios are proposed.  First, dTrans galaxies may be a transitional/intermediate class as dI/dSpiral galaxies transform into dSph/dE galaxies via physical processes in a group environment \citep[e.g.,][]{vanzee04a,vanzee04b}.  Second, it is possible that we are simply observing the natural duty cycle of star formation in isolated or field dI/dSpiral galaxies \citep[e.g.,][]{skillman03,McQuinn15a}.

The NGC~3109 dwarf association provides an opportunity to explore possible scenarios for dTrans formation and evolution, particularly in an environment well isolated from a massive Milky Way-like host galaxy.  A more complete definition of a ``dwarf association" and group membership criteria can be found in \citet{Tully06} and \citet{Kourkchi17}.  In brief, the NGC~3109 association is the closest group of dwarf galaxies \citep[$D\sim 1.4$ Mpc;][]{Tully06} which appear to be physically associated but are not expected to be in dynamical equilibrium.

The NGC~3109 association consists of four dI/dSpiral and two dTrans galaxies.  The four historic members are NGC~3109 (dSpiral; ${\rm M_V} = -14.9$), Sextans~A (dI; ${\rm M_V} = -14.3$), Sextans~B (dI; ${\rm M_V} = -14.5$), and the Antlia dwarf (dTrans; ${\rm M_V} = -10.4$).  We adopt the morphological classifications from \citet{Weisz11} and absolute magnitudes from \citet{McConnachie12}.  \citet{McQuinn15b} has suggested that Leo~P (dI; ${\rm M_V} = -9.3$, $D = 1.6$ Mpc) is also likely a member of the association given the similar distance and spatial proximity to the other dwarf galaxies.  Antlia~B (${\rm M_V} = -9.4$; Table~\ref{table:properties}) has very similar properties to the Antlia dwarf (e.g., smooth spatial distribution of old stars; gas-rich but no H$\alpha$ emission indicating a lack of very recent/ongoing star formation) and so we classify it as dTrans as well.  In total, the association spans a wide absolute magnitude range spread over a large projected area ($\sim 1$ Mpc projected spatial extent of the group).

Figures~\ref{fig:all_sfh}c and \ref{fig:all_sfh}d show the {\it HST} SFHs for all members of the association: NGC~3109, Antlia \citep{Weisz11}, Sextans~A, Sextans~B \citep{Weisz14}, Leo~P \citep{McQuinn15b}, and Antlia~B (this work).  We also show the mean SFH for dI galaxies ($N=8$) and dTrans ($N=5$) in the Local Group \citep[from][]{Weisz14} as dotted lines.  The confidence regions of the mean dI and dTrans SFHs (shown as gray bands) are the standard error in the mean.  Taken together, the NGC~3109 association of dwarfs shows SFHs consistent with the broader dI/dTrans population of the Local Group.  If we compare the SFHs of the NGC~3109 association dwarfs to the sample of field dI/dTrans galaxies from \citet[their Fig. 11]{Weisz14}, we find excellent agreement both in the mean SFH and in the overall spread.  There is likely some overlap in samples (i.e., NGC~3109 association galaxies were likely included in the \citealt{Weisz14} sample of isolated galaxies), but the \citet{Weisz14} sample is clearly larger.

The more luminous members of the NGC~3109 association (NGC~3109, Sextans~A, Sextans~B) show a slightly more rapid growth in mass than the low-mass systems at early times ($t \gtrsim 9$ Gyr ago), perhaps consistent with the expectation that they were born in more massive dark matter halos at early times.  Although they appear to track nicely with the mean SFH for dTrans galaxies until $\sim 3$ Gyr ago, we note that their general properties are more consistent with dI galaxies.

The three lowest luminosity systems (Antlia, Antlia~B, Leo~P) show very similar star formation histories, particularly at early times.  Consistent with their dTrans classification, the SFHs show that Antlia and Antlia~B have formed 95\% or more of their stars in the first $\sim 10$ Gyr and track nicely with the mean dTrans SFH within the last $\sim 3$ Gyr (Figure~\ref{fig:all_sfh}d).  Additionally, both Antlia and Antlia~B have a \mhi/$M_*$ ratio consistent with other Local Volume dwarf galaxies of similar size \citep[see Fig.3 in][]{Sand15b}, despite the fact that they have no active star formation like the typical dI galaxy.  Leo~P, however, shows evidence for active star formation \citep[e.g., O-star embedded in an \ion{H}{2} region;][]{Skillman13, Evans19} consistent with a dI morphological classification.  

The spatial proximity of Antlia and Antlia~B to NGC~3109 suggests that their nature as dTrans galaxies is due to environmental influences.  The Antlia dwarf in particular shows clear evidence of tidal disturbance both in the stellar and gas components.  \citet{Penny12} have shown that Antlia displays stellar tidal tails that are likely the result of an interaction with NGC~3109 (approximately 1 Gyr ago) that may have resulted in the asymmetric \ion{H}{1} warp in NGC~3109 \citep{Barnes01b}.  In addition, \ion{H}{1} gas in Antlia is offset from the main body of the galaxy by $\sim 1\arcmin$ and aligned with the northwest extension of the stellar tidal tail \citep{Ott12}.  

Although Antlia~B shows no evidence of stellar tidal distortion, the relatively close spatial proximity to NGC~3109 suggests that its dTrans properties may also be environmentally driven.  The Green Bank Telescope observations of Antlia~B described in \citet{Sand15b} were optimized for high-sensitivity detection of \ion{H}{1}, but the large beam size ($\sim 9\arcmin$) relative to the small half-light radius of Antlia~B means that no spatial information about the \ion{H}{1} is available.  A future study of Antlia~B using high resolution VLA observations will explore the gas kinematics and distribution in more depth, allowing us to test whether the dTrans properties of this galaxy are environmentally-driven.

\section{Conclusions}\label{sec:conclusions}

\begin{figure*}
\begin{center}
\hspace*{-2cm}\includegraphics[scale=0.6,angle=0]{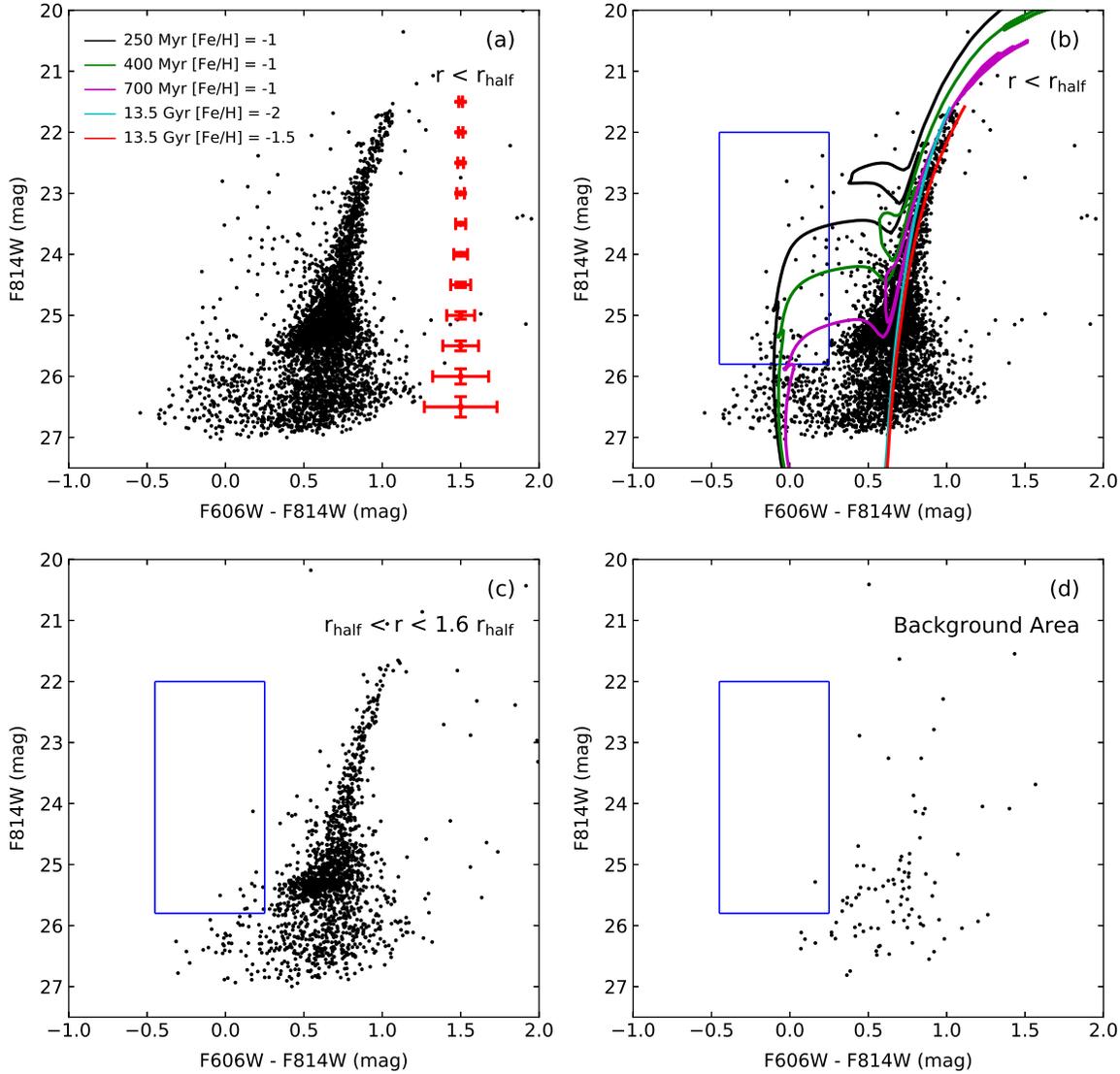}
\caption{Optical CMDs of Antlia~B.  (a)  Point sources within the half-light radius of Antlia~B are shown in black.  The median color (F606W-F814W) and F814W magnitude errors are shown as a function of magnitude as red points.  (b) Same as panel (a) but showing  theoretical isochrones for a range of stellar metallicities and ages (see Section~\ref{sec:cmd}). The blue box highlights the young, blue stellar populations associated with Antlia~B.  Panels (c) and (d) show the CMD in equal area regions outside of the half-light radius of Antlia~B (see Fig.~\ref{fig:props}).  As indicated by the blue selection box, the young, blue stellar populations show a centrally concentrated spatial distribution (see also Figure~\ref{fig:props}).
\label{fig:cmd}}
\end{center}
\end{figure*}

\begin{figure*}
\begin{center}

\includegraphics[width=1.0\textwidth,trim={0cm 0cm 0cm 0cm},clip]{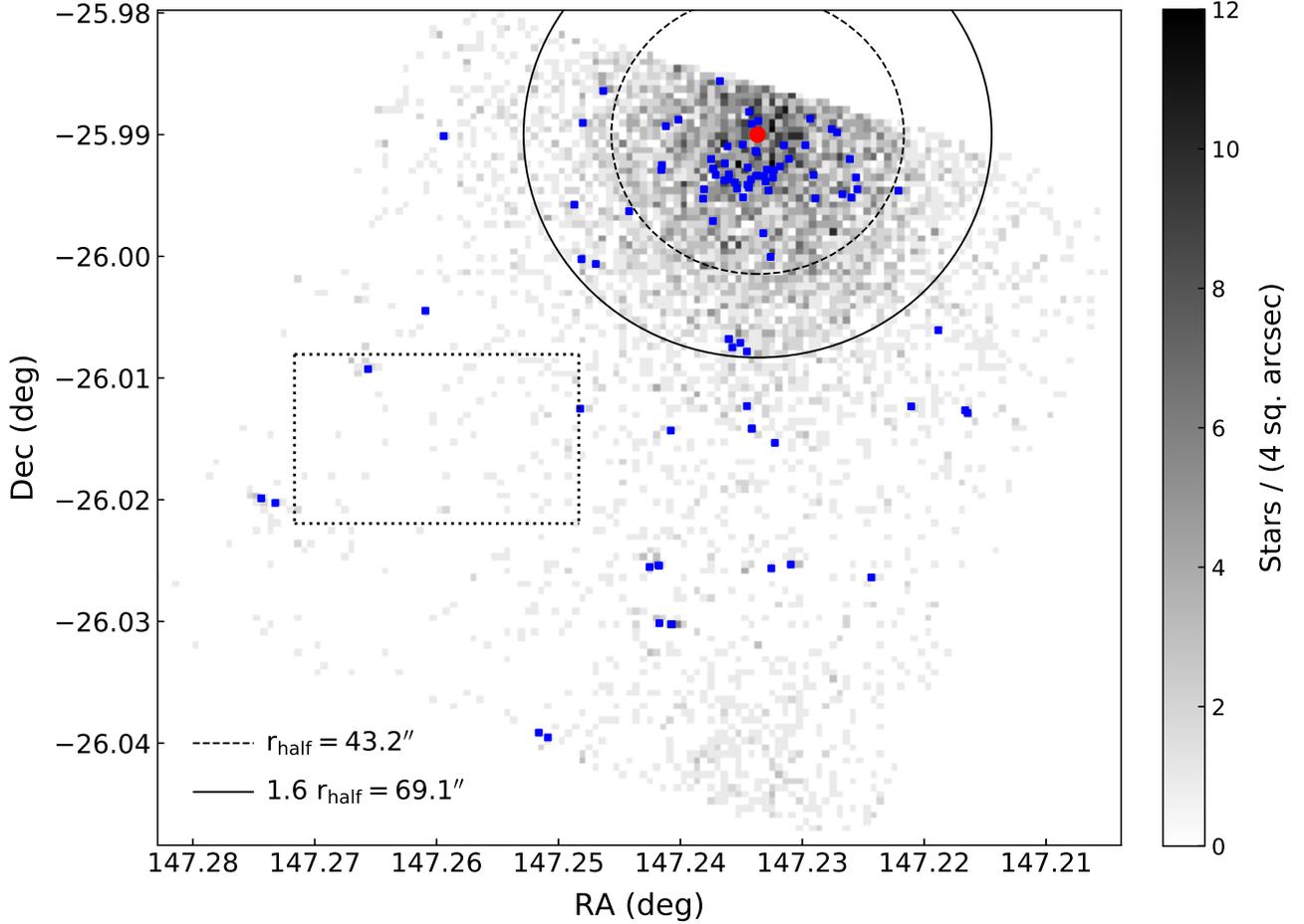}

\caption{Spatial distribution of the Antlia~B stellar populations.  The greyscale bins show the surface density of all stars brighter than the 90\% completeness limit (F814W $< 26.3$; see Figure~\ref{fig:cmd}).  The blue points denote the individual young, blue stars in the selection box shown in Figure~\ref{fig:cmd}. The younger stellar populations show a strong central concentration relative to the more extended old, metal-poor RGB and RC populations.  The red dot denotes the galaxy center as measured in the ground-based imaging \citep{Sand15a}.  The dashed ellipse is drawn with the semi-major axis equal to the half-light radius as measured by \citet[$r_{\rm half} = 43.2\arcsec = 273~\rm{pc}$, ellipticity = 0.3, position angle $= 4~{\rm degrees}$]{Sand15a}.  An annular region between 1 and $1.6~r_{\rm half}$ ({\it solid line}) encloses an area identical to that within $r_{\rm half}$, taking into account the missing area off the HST pointing.  The dotted square region (centered at $r = 3.2~r_{\rm half}$) shows a random background region of identical area for comparison.  The CMDs of stars within these three regions are shown in Figure~\ref{fig:cmd}.
\label{fig:props}}
\end{center}
\end{figure*}

\begin{figure*}
\begin{center}

\includegraphics[scale=.60,angle=0]{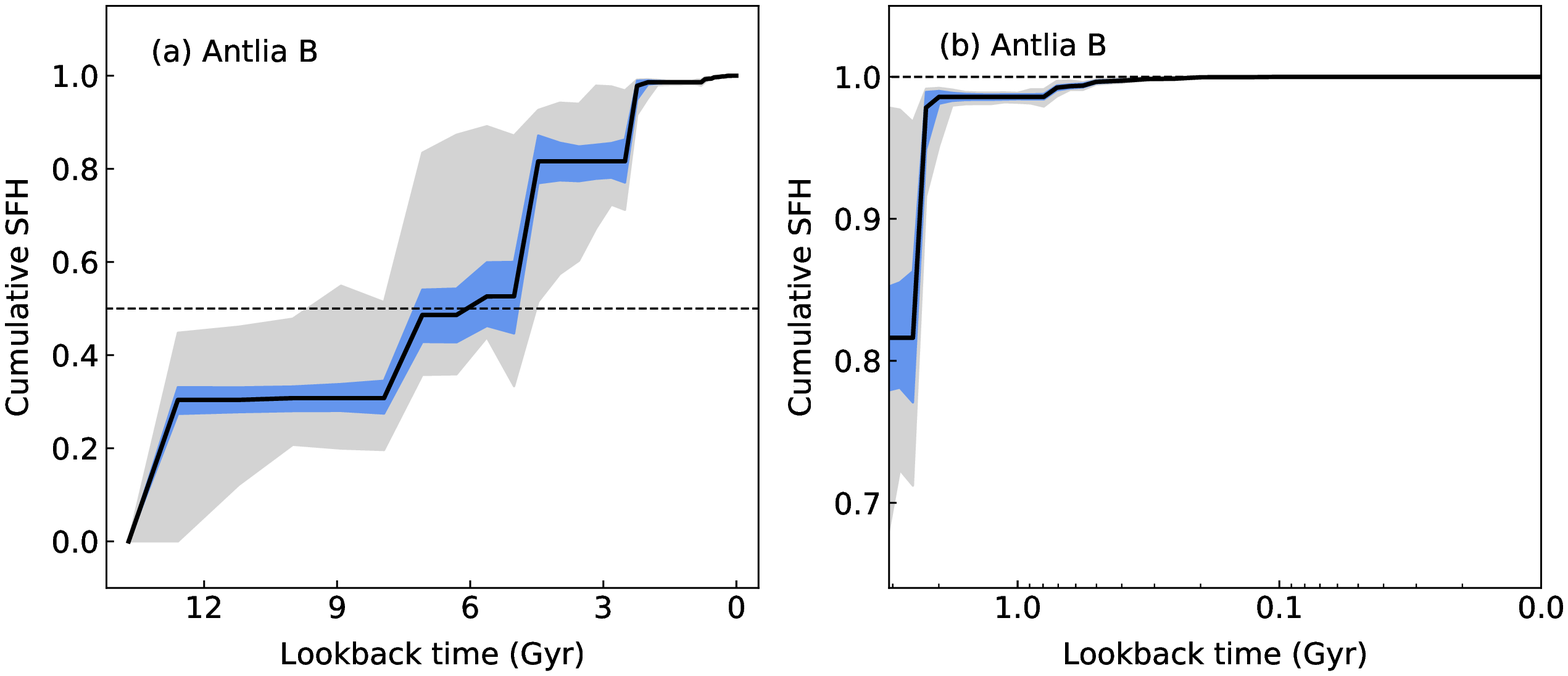}
\includegraphics[width=1.0\textwidth,trim={0cm 0cm 0cm 0cm},clip]{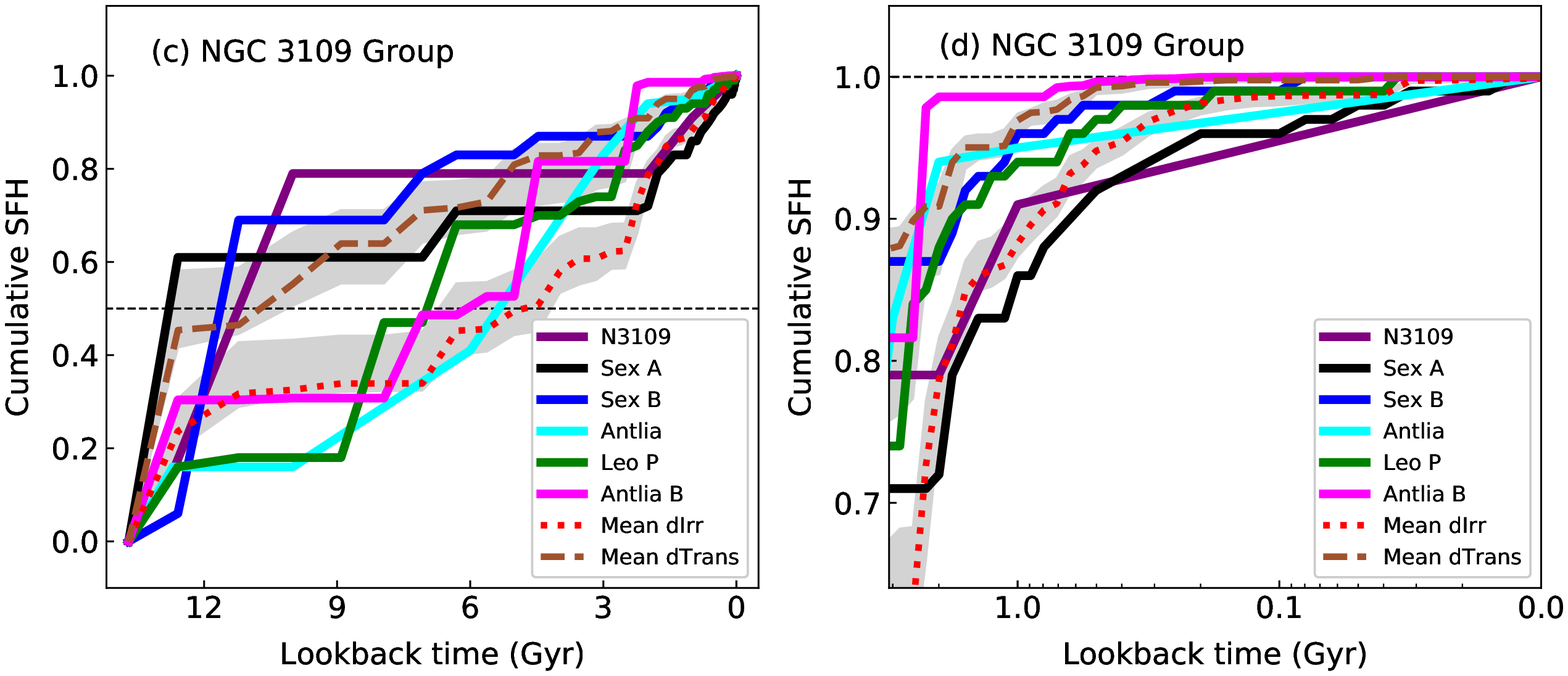}
\caption{Cumulative SFH for Antlia~B (top panels) and the cumulative SFHs for other members of the NGC~3109 dwarf association (bottom panels) taken from the literature.  See Section~\ref{sec:sfh} for a description of the analysis and Section~\ref{sec:discussion} for a discussion and relevant references for individual galaxies.  The left panels (a, c) show the complete SFH while the right panels (b, d) highlight the recent ($t < 3$ Gyr) SFH.  For Antlia~B, the blue shading shows the $16\%/84\%$ confidence regions when accounting for random errors only, while the gray region shows the same confidence intervals when including systematic errors.  The sources of random and systematic uncertainty are described in Section~\ref{sec:sfh}.  For the NGC~3109 group, we show the mean dI and mean dTrans SFHs from \citet{Weisz14}.  The gray confidence regions on the mean SFHs reflect the standard error in the mean.
\label{fig:all_sfh}}
\end{center}
\end{figure*}

In this paper, we present deep {\it HST} imaging of the gas-rich, faint dwarf galaxy Antlia~B discovered as part of our wide-field imaging survey for satellites of nearby low mass host galaxies. Our primary results are as follows:

\begin{enumerate}

    \item We obtain a refined TRGB distance of $D = 1.35 \pm 0.06$ Mpc ($(m-M)_{\rm 0} = 25.65 \pm 0.10$) using the {\it HST} data (see Section~\ref{sec:distance}).  This is slightly more distant than but consistent with the ground-based TRGB determination by \citet{Sand15a}.  Given the distance and projected separation ($\sim 70$ kpc), Antlia~B is clearly a member of the NGC~3109 dwarf association \citep{Tully06,Kourkchi17}.

    \item The CMD of Antlia~B shows both an old, metal-poor stellar population and a small population of young, more metal-rich stars with ages $\lesssim 1$ Gyr (see Figure~\ref{fig:cmd}).  Consistent with previous ground-based imaging from \citet{Sand15a}, we find no evidence for very recent star formation ($\sim 10-100$ Myr timescales).  The young, blue population of stars in Antlia~B are spatially concentrated towards the galaxy center (see Figure~\ref{fig:props}).
    
    \item We derive the SFH of Antlia~B (see Section~\ref{sec:sfh}) using the {\tt MATCH} algorithm \citep{Dolphin02} following the methodology of \citet{Weisz11,Weisz14}.  The SFH is shown in Figure~\ref{fig:all_sfh} and is consistent with the typical dI/dTrans galaxy in the Local Group.  Antlia~B shows a slow, constant growth in mass at early times (first $\sim10-11$ Gyr).  Consistent with a dTrans galaxy classification (see Section~\ref{sec:discussion}), Antlia~B has had very little star formation since this time despite being relatively gas-rich (\mhi $\sim 3\times 10^5 {\rm M}_\odot$; \citealt{Sand15a}).  The SFH indicates that only $\sim 1\%$ of Antlia~B's mass formed in the last $\sim 2-3$ Gyr.
    
    \item All members of the NGC~3109 dwarf association have {\it HST}-derived SFHs and we present a systematic comparison in Section~\ref{sec:discussion}.  All six dwarf galaxies -- NGC~3109, Sextans~A, Sextans~B, Antlia, Antlia~B, and Leo~P -- show SFHs consistent with the mean dI/dTrans population in the Local Group (see Figure~\ref{fig:all_sfh}), particularly when considering the isolated field sample of \citet{Weisz14}.  
    
\end{enumerate}

Both Antlia~B and Antlia are likely satellites of NGC~3109 as suggested by their spatial proximity, linear distances, and heliocentric systemic \ion{H}{1} velocities.  Despite the isolated and low density environment of the NGC~3109 association, the evidence for dynamical interactions between Antlia and NGC~3109 suggests that the suppression of star formation may occur even around very low mass primary hosts like NGC~3109 \citep[stellar mass $\sim 8 \times 10^7~{\rm M}_\odot$;][]{McConnachie12}.

Similar studies to this -- which combine uniform, detailed SFH studies based on a spatially-complete imaging survey -- are currently rare in dwarf association/group environments but are essential for building a more complete picture of dwarf galaxy evolution.  For example, given the small numbers of galaxies in the NGC~3109 group, finding weak trends in SFHs with properties like luminosity, morphology, and/or radial distance from a primary host are inevitable.  More studies of dwarf associations will not only provide insights into possible correlations but will allow for broader studies into the role of environment in shaping the SFHs of the lowest luminosity dwarf satellites.

\acknowledgments
 
 JRH acknowledges support from HST award GO-14078 and the hospitality of the Texas Tech University Department of Physics and Astronomy and the University of Arizona Department of Astronomy/Steward Observatory.  SMA is supported by the National Science Foundation Graduate Research Fellowship under Grant DGE 1752814.  Research by DJS is supported by NSF grants AST-1821967, 1821987, 1813708 and 1813466. Research by DC is supported by NSF grant AST-1814208, and by NASA through grants number HST-GO-15426.007-A and HST-GO-HST-GO-15332.004-A from the Space Telescope Science Institute, which is operated by AURA, Inc., under NASA contract NAS 5-26555. KS acknowledges support from the Natural Sciences and Engineering Research Council of Canada.  B.W. and J.C were supported by an NSF Faculty Early Career Development (CAREER) award (AST-1151462). DRW acknowledges fellowships from the Alfred P. Sloan Foundation and the Alexander von Humboldt Foundation. AHGP is supported by NSF grant AST-1813628. This work was partially performed at the Aspen Center for Physics, which is supported by National Science Foundation grant PHY-1607611. 

\vspace{5mm}
\facilities{HST}

\software{
astropy \citep{astropy13,astropy18}, \texttt{MATCH} \citep{Dolphin02}   
        }

\bibliographystyle{aasjournal}

\clearpage

\begin{deluxetable*}{lll}
\tablecolumns{3}
\tablecaption{Properties of Antlia~B\label{table:properties}}
\tablehead{
\colhead{Parameter}  & \colhead{Value} & \colhead{Source} \\
}
\startdata
RA$_{0}$ (h:m:s) & 09:48:56.08 $\pm$2.1" & \citet{Sand15a}\\
DEC$_{0}$ (d:m:s) & -25:59:24.0 $\pm$3.8" & \citet{Sand15a}\\
$(m-M)_{\rm 0}$ (mag) & 25.65 $\pm$0.10 & This work\\
$D$ (Mpc) & 1.35$\pm$0.06 & This work\\
$D_{\rm proj}~{\rm (kpc)}$ & $73$ & -- \\
$M_{V}$ (mag) & $-$9.7$\pm$0.6 & \citet{Sand15a} \\
$r_{\rm half}$ (arcsec) & 43.2$\pm$4.2 & \citet{Sand15a}\\
$r_{\rm half}$ (pc) & 273$\pm$29 & \citet{Sand15a}\\
$\epsilon$ & 0.30$\pm$0.05 & \citet{Sand15a}\\
$\theta$ (deg) & 4.0$\pm$12.0 & \citet{Sand15a}\\
$S_{21}$ (Jy km s$^{-1}$) & 0.72$\pm$0.05 & \citet{Sand15a}\\
$W50_{HI}$ (km s$^{-1}$) & 17$\pm$ 4 & \citet{Sand15a}\\
$M_{HI}$ (10$^5$ $M_{\odot}$) & 2.8 $\pm$ 0.2 & \citet{Sand15a}\\
$v_{helio, HI}$ (km s$^{-1}$)& 376 $\pm$ 2 & \citet{Sand15a}
\enddata
\end{deluxetable*}

\begin{splitdeluxetable*}{ccccccBcccccccccBcccccccc}
\tabletypesize{\scriptsize}
\tablecaption{Photometry of Resolved Stars in the HST/ACS imaging of Antlia~B \label{tab:photometry}}
\tablehead{
\colhead{\#}& \colhead{$\alpha$ (2000)} & \colhead{$\delta$ (2000)} & \colhead{X} & \colhead{Y} & \colhead{Object Type} & \colhead{F606W} & \colhead{$\sigma_{\rm F606W}$} & \colhead{${\rm SNR}_{\rm F606W}$} & \colhead{${\rm Sharp}_{\rm F606W}$} & \colhead{${\rm Round}_{\rm F606W}$} & \colhead{${\rm Crowd}_{\rm F606W}$} & \colhead{${\rm Flag}_{\rm F606W}$} & \colhead{F814W} & \colhead{$\sigma_{\rm F814W}$} & \colhead{${\rm SNR}_{\rm F814W}$} & \colhead{${\rm Sharp}_{\rm F814W}$} & \colhead{${\rm Round}_{\rm F814W}$} & \colhead{${\rm Crowd}_{\rm F814W}$} & \colhead{${\rm Flag}_{\rm F814W}$} & \colhead{E(B-V)} & \colhead{${\rm A}_{\rm F606W}$} & \colhead{${\rm A}_{\rm F814W}$} \\
\colhead{}& \colhead{(deg)}& \colhead{(deg)}& \colhead{pix}   & \colhead{pix}   & \colhead{} & \colhead{(mag)} & \colhead{(mag)} & \colhead{} & \colhead{} & \colhead{} & \colhead{} & \colhead{} & \colhead{(mag)} & \colhead{(mag)} & \colhead{} & \colhead{} & \colhead{} & \colhead{} & \colhead{} & \colhead{(mag)} & \colhead{(mag)} & \colhead{(mag)}
} 
\colnumbers
\startdata 
75 & 147.2344204 & -26.0254116 & 2945.5 & 1464.66 & 1.0 & 18.875 & 0.011 & 97.8 & -0.06 & 0.134 & 0.226 & 0.0 & 17.694 & 0.013 & 85.3 & -0.256 & 0.162 & 0.168 & 0.0 & 0.081 & 0.200 & 0.123 \\
79 & 147.2284601 & -26.0202965 & 3181.9 & 1942.64 & 1.0 & 19.193 & 0.003 & 388.6 & -0.065 & -0.003 & 0.0 & 0.0 & 17.658 & 0.004 & 286.2 & -0.139 & 0.091 & 0.0 & 0.0 & 0.080 & 0.199 & 0.123  \\
82 & 147.2344531 & -26.0254638 & 2944.8 & 1460.4 & 1.0 & 18.907 & 0.019 & 56.8 & -0.014 & 0.12 & 0.403 & 0.0 & 17.875 & 0.022 & 50.0 & -0.208 & 0.227 & 0.394 & 0.0 & 0.081 & 0.200 & 0.123 \\
92 & 147.2510956 & -26.0251222 & 1924.61 & 1115.11 & 1.0 & 19.346 & 0.002 & 706.4 & -0.01 & 0.011 & 0.018 & 0.0 & 17.973 & 0.001 & 862.0 & -0.012 & 0.008 & 0.02 & 0.0 & 0.081 & 0.200 & 0.123 \\
93 & 147.2167233 & -26.0185591 & 3852.73 & 2319.9 & 1.0 & 19.763 & 0.003 & 427.2 & 0.019 & 0.002 & 0.019 & 0.0 & 17.74 & 0.004 & 268.3 & -0.005 & 0.032 & 0.023 & 2.0 & 0.080 & 0.199 & 0.123 \\
94 & 147.2555204 & -25.9930577 & 865.52 & 3186.38 & 1.0 & 19.028 & 0.001 & 800.6 & -0.002 & 0.015 & 0.047 & 0.0 & 18.421 & 0.002 & 696.9 & 0.024 & 0.007 & 0.04 & 2.0 & 0.080 & 0.198 & 0.122 \\
95 & 147.2436497 & -26.0350494 & 2621.83 & 608.33 & 1.0 & 19.729 & 0.002 & 462.6 & -0.009 & 0.011 & 0.018 & 2.0 & 18.013 & 0.003 & 402.5 & -0.072 & 0.076 & 0.025 & 0.0 & 0.081 & 0.200 & 0.124 \\
103 & 147.281446 & -26.0271178 & 128.75 & 307.92 & 1.0 & 20.333 & 0.003 & 406.9 & 0.006 & 0.021 & 0.037 & 0.0 & 17.976 & 0.001 & 945.8 & -0.004 & -0.002 & 0.043 & 0.0 & 0.081 & 0.200 & 0.123 \\
112 & 147.222213 & -26.0191429 & 3533.28 & 2158.92 & 1.0 & 19.918 & 0.003 & 399.0 & -0.004 & 0.017 & 0.031 & 1.0 & 18.538 & 0.002 & 720.1 & 0.0 & -0.002 & 0.022 & 1.0 & 0.080 & 0.199 & 0.123 \\
114 & 147.231019 & -26.0430296 & 3586.19 & 347.96 & 1.0 & 20.065 & 0.002 & 481.4 & -0.005 & 0.02 & 0.021 & 0.0 & 18.51 & 0.001 & 755.3 & 0.001 & 0.007 & 0.024 & 0.0 & 0.081 & 0.201 & 0.124 \\

\enddata
\tablecomments{Photometric catalog of resolved stars in the {\it HST/ACS} data set used in this study.  Sources which did not pass the point source selection criteria described in Section~\ref{sec:datareduce} were not included in this catalog.  For completeness, we provide the full output from our {\tt DOLPHOT} photometry and refer the reader to the {\tt DOLPHOT} documentation for specific details on the column descriptions.  Note that the photometry in this table is not corrected for Milky Way extinction, but extinction values are provided for convenience.  All figures in this paper have been correct for Galactic extinction using Columns (21)-(23) described below.  This table is available in its entirety in a machine-readable form online. A small portion of the data is shown as an example of the form and content of the table.
\\
\\
(1) Object identification number. (2)-(5) Position of sources in the celestial equatorial and image frames of reference. (6) Source object type as described by {\tt DOLPHOT}. (7)-(13) Calibrated magnitudes, errors, signal-to-noise (SNR), shape parameters (sharp, round), crowding parameter, and quality flag for the F606W photometry.  (14)-(2)  Same as for columns (7)-(13) but for the F814W photometry.  (21)  Color excess for each source from \citet{Schlegel98} dust maps. (22)-(23)  Milky Way extinction values in F606W and F814W filters derived using the coefficients from \citet{Schlafly11}.}
\end{splitdeluxetable*}

\begin{deluxetable*}{cccc}
\tablecaption{Cumulative Star Formation History of Antlia~B\label{tab:sfh}}

\tablehead{
\colhead{log($t$)  [yr]} & \colhead{$f$} & \colhead{${\sigma_{\rm ran}}$$(84\%, 16\%)$} & \colhead{${\sigma_{\rm tot}}$$(84\%, 16\%)$} \\
\colhead{(1)} & \colhead{(2)} & \colhead{(3)} & \colhead{(4)}
}
\startdata
\input{tmp2.out}
\enddata
\tablecomments{Cumulative star formation history (SFH) for Antlia~B (see Section~\ref{sec:sfh}). (1) Epoch over which the fractional stellar mass growth $f$ is calculated.  (2) Fraction of the total stellar mass formed prior to the corresponding epoch.  (3) Upper and lower random uncertainties on the fractional stellar mass.  (4) Upper and lower total uncertainties (random and systematic) on the fractional stellar mass.  \\
\\
This table is available in its entirety in a machine-readable form online. A small portion of the data is shown as an example
of the form and content of the table.
}
\end{deluxetable*}

\end{document}